\documentclass[review]{elsarticle}

\usepackage{lineno,hyperref}
\usepackage{amsmath}
\modulolinenumbers[5]

\journal{Chemical Physics Letters}

\begin{document}

\begin{frontmatter}

\title{On the hierarchical parallelization of \textit{ab initio} simulations}

\author[mymainaddress,mysecondaryaddress]{Sergi Ruiz-Barragan}
\author[mysecondaryaddress]{Kazuya Ishimura}

\author[mymainaddress]{Motoyuki Shiga\corref{mycorrespondingauthor}}
\cortext[mycorrespondingauthor]{Corresponding author}
\ead{shiga.motoyuki@jaea.go.jp}

\address[mymainaddress]{CCSE, Japan Atomic Energy Agency, 178-4-4, Wakashiba, Kashiwa, Chiba, 277-0871, Japan}
\address[mysecondaryaddress]{Department of Theoretical and Computational Molecular Science, Institute for Molecular Science, Myodaiji, Okazaki, Aichi 444-8585, Japan}

\begin{abstract}
A hierarchical parallelization has been implemented 
in a new unified code PIMD-SMASH for \textit{ab initio} simulation where the replicas and the Born-Oppenheimer forces are parallelized.
 It is demonstrated that \textit{ab
initio} path integral molecular dynamics simulations can be carried out very
efficiently for systems up to a few tens of water molecules.
 The code was then used to study a Diels-Alder reaction
 of cyclopentadiene and butenone by \textit{ab initio} string method.
 A reduction in the reaction energy barrier is found in the presence
 of hydrogen-bonded water, in accordance with experiment.
\end{abstract}


\end{frontmatter}


\section{Introduction}
In recent years, the massively parallel architecture has played a key
 role in the advances of modern supercomputers, such as
 the K computer \cite{K-computer}.
In computational chemistry, much effort has been devoted toward
 the development of theoretical methods and numerical algorithms
 to make best use of their parallel performance.
Among them, there is a class of molecular simulations where the
 Born-Oppenheimer (BO) energies and the BO forces can be evaluated
 in parallel for a set of molecular structures of the same system,
 which are referred to as the ``replicas'', ``beads'', or ``images''
 depending on the method used.
This class of molecular simulations is employed for many purposes,
 such as quantum statistics
 (e.g.\ path integral molecular dynamics (path integral MD)
 \cite{tuckerman1993efficient}),
 statistical sampling
 (e.g.\ replica exchange MD \cite{sugita1999replica}),
 nonadiabatic dynamics
 (e.g.\ surface hopping dynamics
 \cite{SurfaceHopping-1}),
 reaction paths
 (e.g.\ Onsager-Machlup method \cite{fujisaki2010onsager},
 string method \cite{e2002string,STRING-method-1})
 or free energy exploration methods
 (e.g.\ constrained MD \cite{carter1989constrained},
 multiple walker metadynamics \cite{CPMD-metadynamics-1,raiteri2006efficient}).\\
Although many of the above methods have been established for empirical force fields, they are expected to be replaced by BO forces based on \textit{ab initio} or first-principles electronic structure calculations, for accurate theoretical prediction.
For this aim, a fortran-based interface code called PIMD\cite{PIMD-code} was originally  developed in our research group for \textit{ab initio} path integral simulations
\cite{PIMD-shiga-1}, see reference \cite{PIMD-Gaussian-1} for a short explanation
 of the code.
The latest version of PIMD supports other replica simlations such as the string method,
 multiple-walker metadynamics or replica exchange methods.
The PIMD code has a subroutine to extract the BO forces from other softwares,
 such as Turbomole\cite{TURBOMOLE}, Mopac, Gaussian\cite{Gaussian,PIMD-Gaussian-1},
 etc., using the intrinsic subroutine `system' that passes unix commands to the shell
 and reading the resulting data from the hard disk.
For the similar purpose, a python-based interface i-Pi code\cite{i-PI-1} was released
 to carry out path integral simulations extracting BO forces from
 FHI-aims \cite{FHI-aims-1} and CP2K \cite{CP2K-1}.
For an efficient \textit{ab initio} simulations with many replicas,
 hierarchical parallelization in the replicas (``replica-parallel'') and
 in the BO forces (``force-parallel'') should be required.
Actually, this function has been employed in CPMD code based on the
 plane-wave density functional theory \cite{CPMD-2}.
However, calling a set of independent
 parallel force calculations via `system' subroutine
 as done in the PIMD code
 is often prohibited in public facilities,
 such as the K computer, for administrative reasons.
Therefore, to enable such a hierarchical parallelization,
 we have made a coherent unification of
 the PIMD code and the \textit{ab initio} electronic structure code,
 Scalable Molecular Analysis Solver for High-performance computing
 systems (SMASH) \cite{SMASH-web}.
The new PIMD-SMASH code is a single MPI program for \textit{ab initio} replica
 simulation based on conventional quantum chemistry with Gaussian basis sets.
In this paper, we illustrate the technical details on how to integrate
 parallel molecular simulation and \textit{ab initio} electronic
 structure codes that have already been parallelized.
We study the advantages and limitations on the parallel performance
 of hierarchical parallelized code for two different simulations,
 i.e.\ path integral MD simulation
 of water clusters and minimum energy path (MEP) calculation
 of Diels-Alder (DA) reaction of cyclopentadiene and butenone
 using the string method.
Finally, scientific discussions have been made on the
 results of the DA reaction which is a typical example of green
 organic synthesis in an aqueous environment.


\section{Methods and implementation}
\begin{figure}[p]
\centering
\caption{a) Schematic of the hierarchical parallelized code PIMD-SMASH. b) Pseudo-code of the unified program PIMD-SMASH in Fortran using MPI subroutines. The comments are marked with ``!". c) Scheme of the MPI processes (numbers), the MPI groups and the MPI communications (coloured rectangle). The red colour indicates the PIMD communication while the blue indicates the SMASH communications of the process '0'. }
\includegraphics[width=0.6\linewidth]{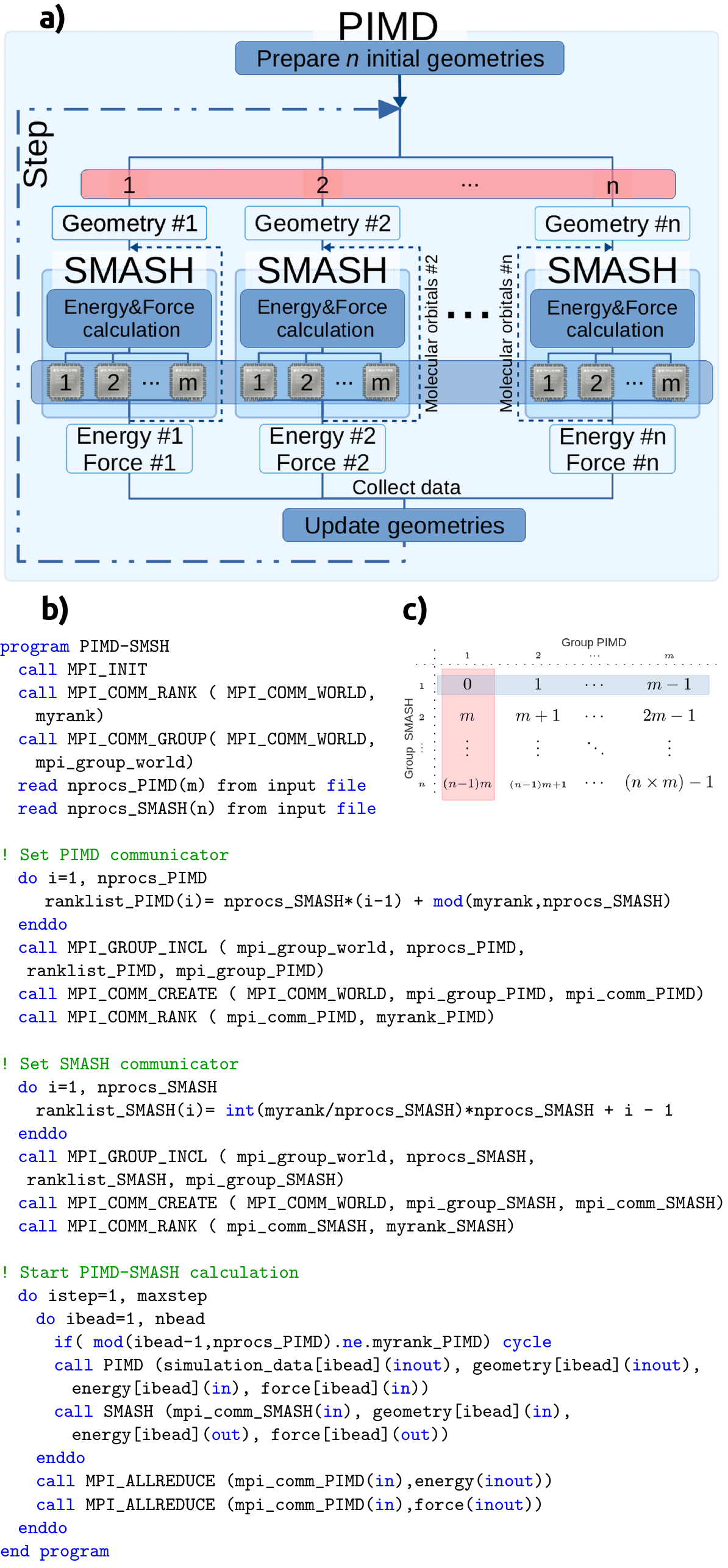}
\label{fig:algorithm}
\end{figure}

The code of SMASH was adapted and introduced in PIMD code to obtain a single program, the PIMD-SMASH.
The PIMD-SMASH code is schematically shown
 in Figure \ref{fig:algorithm}(a).
In PIMD a set of $n$ nuclear configurations (replicas) is generated.
Then, in SMASH the BO energies and forces of the respective replica
 are computed using $m$ cores.
Thus, the hierarchically parallel computation is done
 using $m \times n$ cores in total.
In PIMD the BO energies and forces are gathered, all the positions
 of $n$ replicas are updated according to the simulation method chosen.
Finally, all the replicas are used to start a next step of the MD cycle.


The PIMD-SMASH code uses the MPI library routines, see
 the pseudocode shown in Figure \ref{fig:algorithm}(b).
The hierarchical parallelization requires
 a group of replica-parallel, $mpi\_group\_PIMD$ and
 a group of force-parallel, $mpi\_group\_SMASH$,
 as shown in Figure \ref{fig:algorithm}(c).
Accordingly, the two respective MPI communicators,
 $mpi\_comm\_PIMD$ and $mpi\_comm\_SMASH$,
 are created at the beginning
 of the calculation.
The process ID's for the respective communicators,
 $myrank\_PIMD$ and $myrank\_SMASH$,
 are generated to distribute computational tasks;
$myrank\_PIMD$ is ranged from $0$ to $n-1$ while
 $myrank\_SMASH$ is ranged from $0$ to $m-1$.
The BO energies and forces of all the replicas
 are shared among all the processes.
This is done with the simple sum using
 the $MPI\_ALLREDUCE$ routine,
 since the data is zero wherever not computed.
Using the BO energies and forces, the positions and velocities are updated. The updates are not parallelized thus the improvement in the efficiency will be insignificant because the bottleneck is the force calculation. However, future updates in the code will go in this direction.
To share the information of input files first
 read by the parent process,
 $MPI\_BCAST$ communication is required 
 only once at the beginning of the calculation.
Although SMASH have an initial implementation of hybrid parallelization MPI/OpenMP, this version of PIMD-SMASH does not have this feature available. This option will be introduced in the next version of PIMD-SMASH using the improvement of the next version of SMASH.

Thus, PIMD-SMASH is a seamless MPI code having
 the electronic structure calculations included,
 as opposed to the original version of the PIMD code.
This not only makes the code more efficient
 but also more friendly to both users
 and administrators
(MPI packages have readily accessible documentation and easy tools to manage parallel programs using a job scheduler).
Furthermore, it is easier to extend the functions
 of the code, and to optimize the efficiency.
For instance, the molecular orbitals of
 each replica saved in the memory can be reused
 for the initial guess of the next MD step,
 without hard disk access.\\
We will publish the source code freely
 with GNU General Public License version 3
 \cite{gplv3} in 2016, and a web page is under construction in Japan Atomic Energy Agency (JAEA).
Further details of implementation will be available with the source code. 
The PIMD code facilitates the implementation of user-made forces. In fact, PIMD could also be readily coupled to other electronic structure codes to take advantage of supported
sampling methods.


\section{Results and discussion}
We have completed the coding of PIMD-SMASH in the way described
 in the previous section.
In this section, the parallel performance
 is tested to clarify its advantages and limitations
 on a quantitative basis.
All the calculations were performed on the Fujitsu PRIMERGY BX900 machine
 composed of Intel Xeon X5570 CPUs (8 cores/node)
 in JAEA.


\subsection{Parallel performance: path integral MD of water clusters}

Short runs of \textit{ab initio} path integral MD simulation
 \cite{shiga2001unified} of
 1--64 water molecules (in the linearly aligned configurations)
 were carried out to study the parallel performance.
The electronic structure calculations were
 based on the B3LYP/6-31G$^\ast$ level of theory,
 where 18 basis functions are allocated
 for each water molecule. 
Following the default settings of SMASH, the self-consistent field
 (SCF) cycles with the density threshold of $1.0\times 10^{-6}$
 and the integral cutoff of $1.0\times 10^{-12}$ were used.


%
The parallel efficiency with respect to the beads
 (replicas, in path integral MD simulations)
 is shown in Table \ref{table:pimd-perf1} and \ref{table:pimd-perf2}.
(Note that in the conventional path integral MD simulations
 of water molecules
 at the room temperature, it is known that the number of beads
 required is about 16--64 to converge the average energies
 and distribution functions \cite{shinoda2005quantum}, although
 there are recent efforts to reduce the number of beads
 \cite{PIMD-conv-1,PIMD-conv-2}.)
Table \ref{table:pimd-perf1} shows the average time per step when the number of cores is increased in proportion to the number of beads, i.e. the weak scaling (scaling with fixed system size per processor), while Table \ref{table:pimd-perf2} shows the performance for 64 beads with various number of cores which shows the strong scaling (scaling at fixed system size). 
In both cases, four cores were used for the force calculations such that the results are unaffected by the force-parallel efficiency.
A small but gradual increase of the average time shown in Table \ref{table:pimd-perf1} indicates that the replica-parallel efficiency is high but not perfect.
After extensive tests, it turned out that this deterioration stems from
 the communication overflow when many nodes were occupied by an independent
 set of SMASH calculations.
This problem is heavily machine dependent, maybe particular to BX900, since we
 find the replica-parallel efficiency was nearly perfect when the same
 test was done on other computers.
Even so, the difference of about three seconds between the cases of
 two beads and 64 beads is not very serious.
On the other hand, the strong scaling is above 90\% for the cases
 shown in Table \ref{table:pimd-perf2}.

\begin{figure}[p]
\centering

 \includegraphics[width=0.7\linewidth]{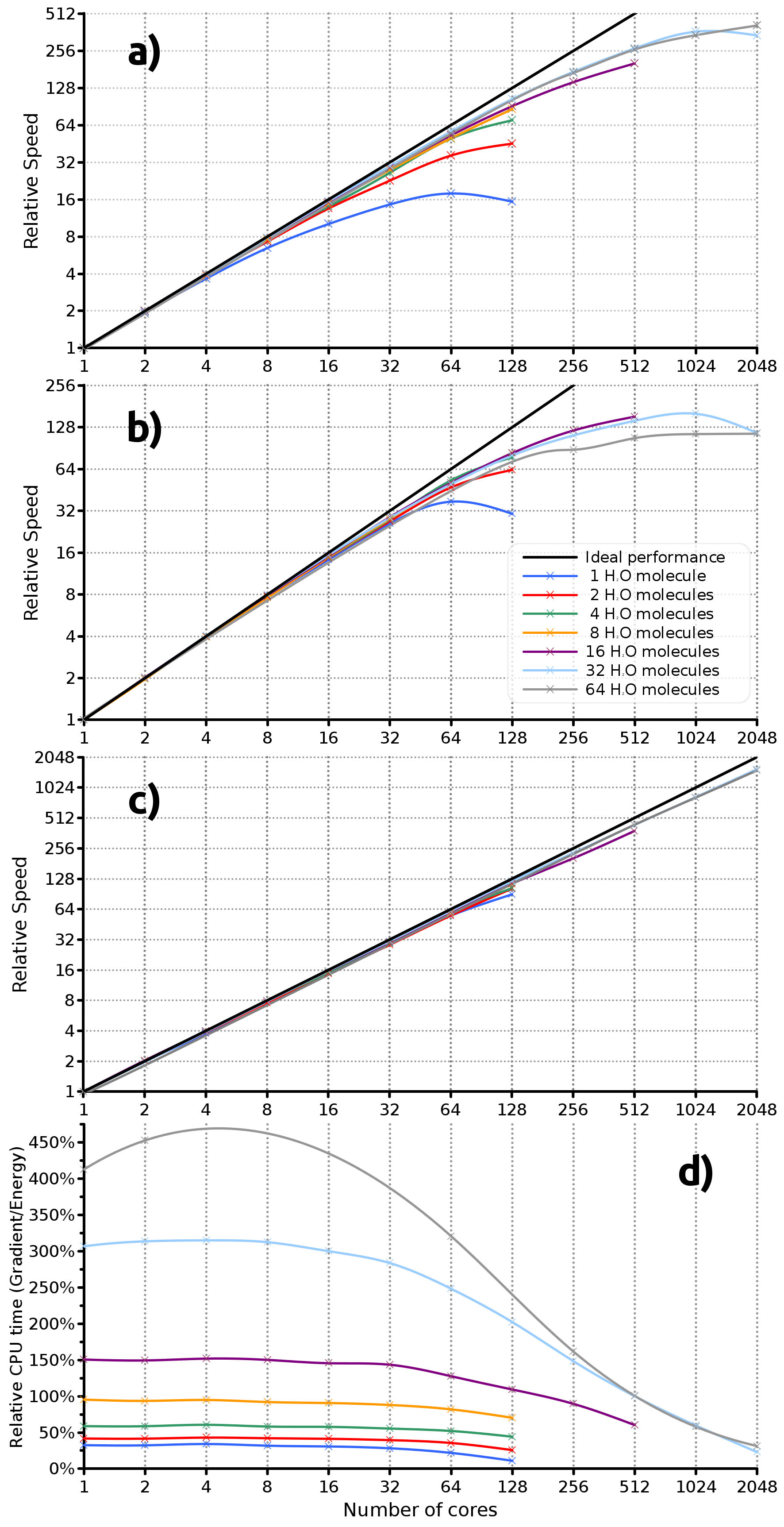}
\caption{
 The parallel performance of path integral MD simulations
 of 1--64 water molecules with two beads.
 (a) The relative speed in the total simulation.
 The relative speed in the calculation of
 (b) the BO energy and (c) the BO force.
 (d) The ratio of the average time spent on
 the BO force and the BO energy.}
 \label{fig:perf}
\end{figure}
\begin{figure}[p]
\centering
\includegraphics[width=1.0\linewidth]{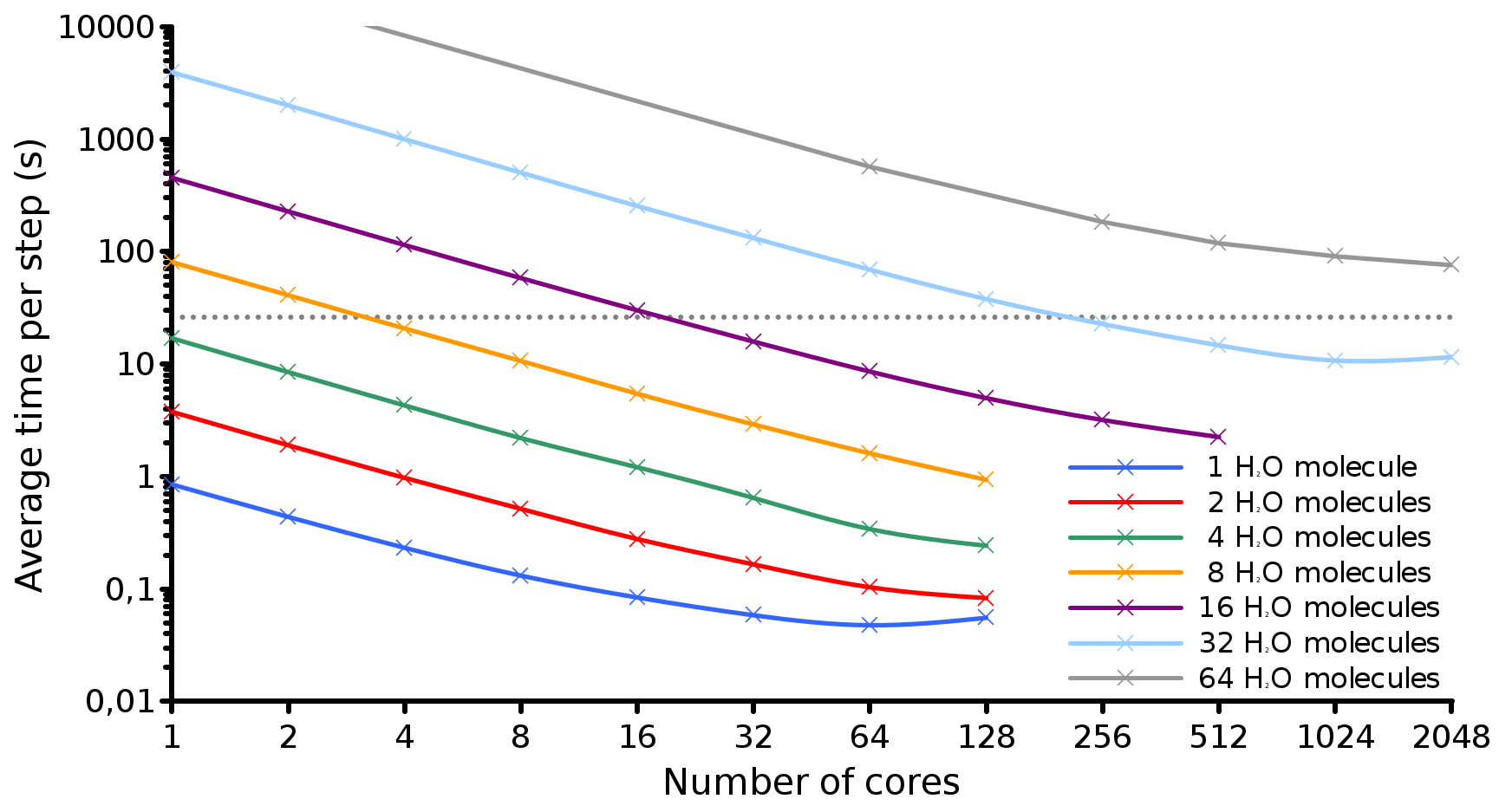}
\caption{Average time per step in
 path integral MD simulations of 1--64 water molecules with two beads.
 The colour codes are the same as those of Figure \ref{fig:perf}.}
\label{fig:aver-times-perf}
\end{figure}

Next, the parallel performance of \textit{ab initio}
 path integral MD simulations were tested
 in the case of two beads.
In Figure \ref{fig:perf}(a), we plot the
 results of the relative speed,
 which is defined as the steps proceeded per unit
 time in comparison with those using a
 single core for the same system.
One can see a general trend that the parallel
 performance is better as the system size
 increases.
For instance, the relative speed for a single water
 molecule stays around 15 when 32 or more cores are used.
This is because the amount of calculation is too small
 to be parallelized efficiently.
On the other hand, the relative speed of 16 water molecules
 grows from 28.7 to 91.1, which amounts to the parallel
 efficiency as high as 71.2\%.\\
To understand the origin of the inefficiency, the relative
 speed is separated between the contributions from
 the BO energies and the BO forces, which are
 shown in Figure \ref{fig:perf}(b) and \ref{fig:perf}(c), respectively.
The force calculation clearly
 outperforms the energy calculation.
The parallel efficiencies in the BO force calculations
 were found to be over 70\% in all the cases studied.
In contrast, the efficiencies are poorer
 in the BO energy calculations, with less
 than 50\% beyond 256 cores.
To see how this feature affects the total timing,
 the ratio of calculation time spent in the BO force
 and the BO energy is displayed in Figure \ref{fig:perf}d.
Here one finds that the ratio is large
 for a system of more than 32 water molecules,
 especially in the case of small number of cores, 1--64.
This explains the fact that the efficiency
 remains high up to 64 cores for large systems.
However, the ratio drops quickly beyond 64 cores
 due to the inefficiency of the energy calculation.
Thus, the total parallel efficiency decreases
 as the average time per step is dominated
 by the BO energy contribution. \\
In Figure \ref{fig:aver-times-perf},
 we show the average time per step in
 path integral MD simulation.
This index is crucial for any kind of MD simulations.
For instance, it is often the case that the (path integral)
 MD trajectory of several tens of picoseconds,
 which amounts to a few hundred thousands of steps,
 is required to obtain statistically converged
 results of the pair distributions of liquid water.
Say, to complete 100000 steps in 30 days of continuous
 run, the average time per step must be less than 25.92 seconds.
This value is marked as a dotted line
 in Figure \ref{fig:aver-times-perf}.
One can see that
 the results are below the line for a single core
 for the small systems up to 4 water molecules.
For 8, 16 and 32 water molecules,
 there is a crossover
 at around 4, 19 and 210 cores,
 respectively.
In this way, as the system size increases the
 number of cores required is `delayed' to a larger value.
Finally, in the case of
 64 water molecules, the crossover is not found
 for any number of cores, which means
 that such a calculation can never be efficiently executed.
This is the reason why the MD simulation of large systems
 remains as a challenging issue.


\subsection{Parallel performance: string method for DA reaction}

\begin{figure}[p]
\centering
\includegraphics[width=1.0\linewidth]{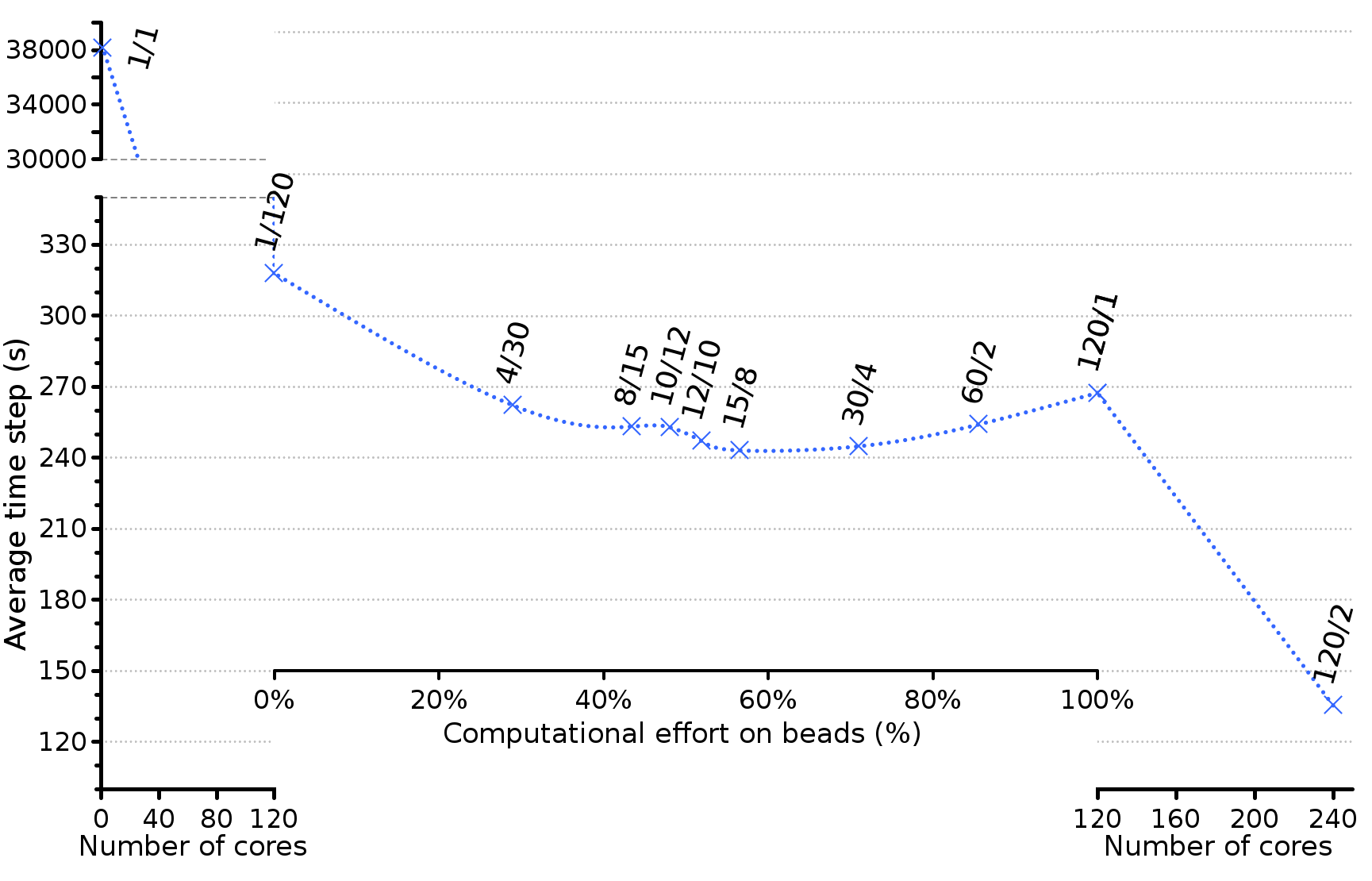}
\caption{Average time per step of the string method
 for a DA reaction of cyclopentadiene and butenone.
 120 images are used to describe the structural change along the string.
 The image parallel number $n$ and the force-parallel number $m$
 are given as ``$n/m$'' where
 $n \times m$ is the total number of cores used.
 The index $\log_{120}^{}(n)$ measures the computational effort put on
 the image parallel number over the force-parallel number
 for a fixed number of cores $n \times m = 120$.}

\label{fig:hier-perf-1}
\end{figure}

Next, hierarchical parallelization is applied to
 the string method for a simulation of
 the DA reaction of cyclopentadiene and butenone.
The string is described in terms of 120 images
 (replicas in the string method)
 that interconnect the reactant minimum to the product minimum,
 which are shown later in Figure \ref{fig:bar-rest}.
Here, the electronic structure calculations are
 based on
 the B3LYP/6-31G$^{\ast\ast}$ level of theory.
The results are summarized in Figure \ref{fig:hier-perf-1}.
Using a single core it took 38159 seconds per step,
 where a step in the string method refers to the force calculation
 of all the images followed by the update of the whole string.
With 120 cores,
 the average times per step reduced to 244.8--318.0 seconds
 depending on how the cores are distributed.
In hierarchical parallelization, the cores can be used
 in different products of the force-parallel number $m$
 and the image-parallel number $n$,
 and the best combination seems to be moderate
 values,
 $n=15$ and $m=8$ in the present case.
Clearly, the image-parallel has better load balancing and it is advantageous in reducing the
 communication during the force calculations,
 but it is also disadvantageous in calculating
 many images at the same time and waiting for
 all to finish.
This aspect is not negligible when the number of
 SCF cycles varies among the images,
 which is the case between equilibrium
 and transition state.
This point is particularly important in the string method
 because the images are usually in very different geometries,
 but it may become less serious in other methods such as
 path integral MD simulations.
Comparing the fully image-parallel cases,
 the time per step of the double force-parallel
 using 240 cores, was almost a half of that of
 the single force-parallel, using 120 cores.
This confirms that the efficiency of the string method
 when the force-parallel number is sufficiently small.


\subsection{Application: Diels-Alder reaction}

Finally, we have analyzed the MEP of the DA reaction of cyclopentadiene and butenone obtained by the string method using hierarchical parallelization.
This reaction is known as one of the earliest organic reactions
 found to be accelerated in aqueous solution \cite{breslow1991hydrophobic}.
Initially, it was proposed that the acceleration is
 due to hydrophobicity \cite{breslow1991hydrophobic}.
More recently, some theoretical works
 \cite{dial-MVK-water-1,dial-MVK-water-2,dial-MVK-water-3,dial-MVK-water-4}
 suggested that the hydrogen bond plays the key role in the acceleration. 
All the MD simulations of
 aqueous solutions done so far are based on the semi-empirical methods
 \cite{dial-MVK-water-2,dial-MVK-water-3,dial-MVK-water-4}
 since the statistical accuracy less than 1 kcal/mol
 would be formidable for \textit{ab initio} MD.
To our knowledge, the only \textit{ab initio} study is
 reference \cite{dial-MVK-water-2} where Hartree-Fock calculations
 of the molecular cluster system with one water molecule
 were completed using the constrained geometry optimization.
Here we present the \textit{ab initio} calculation of the
 molecular cluster system with different numbers
 water molecules (0-2).
For each system, the full MEP from the reactant minimum
 to the product minimum is obtained without
 any geometrical constraints, which allows the estimation
 of the activation energies of forward and backward reactions.
To account for the electron correlation effects,
 the calculation is based on the B3LYP/6-31G$^{**}$
 level of theory.

\begin{figure}[p]
\centering
\includegraphics[width=0.7\linewidth]{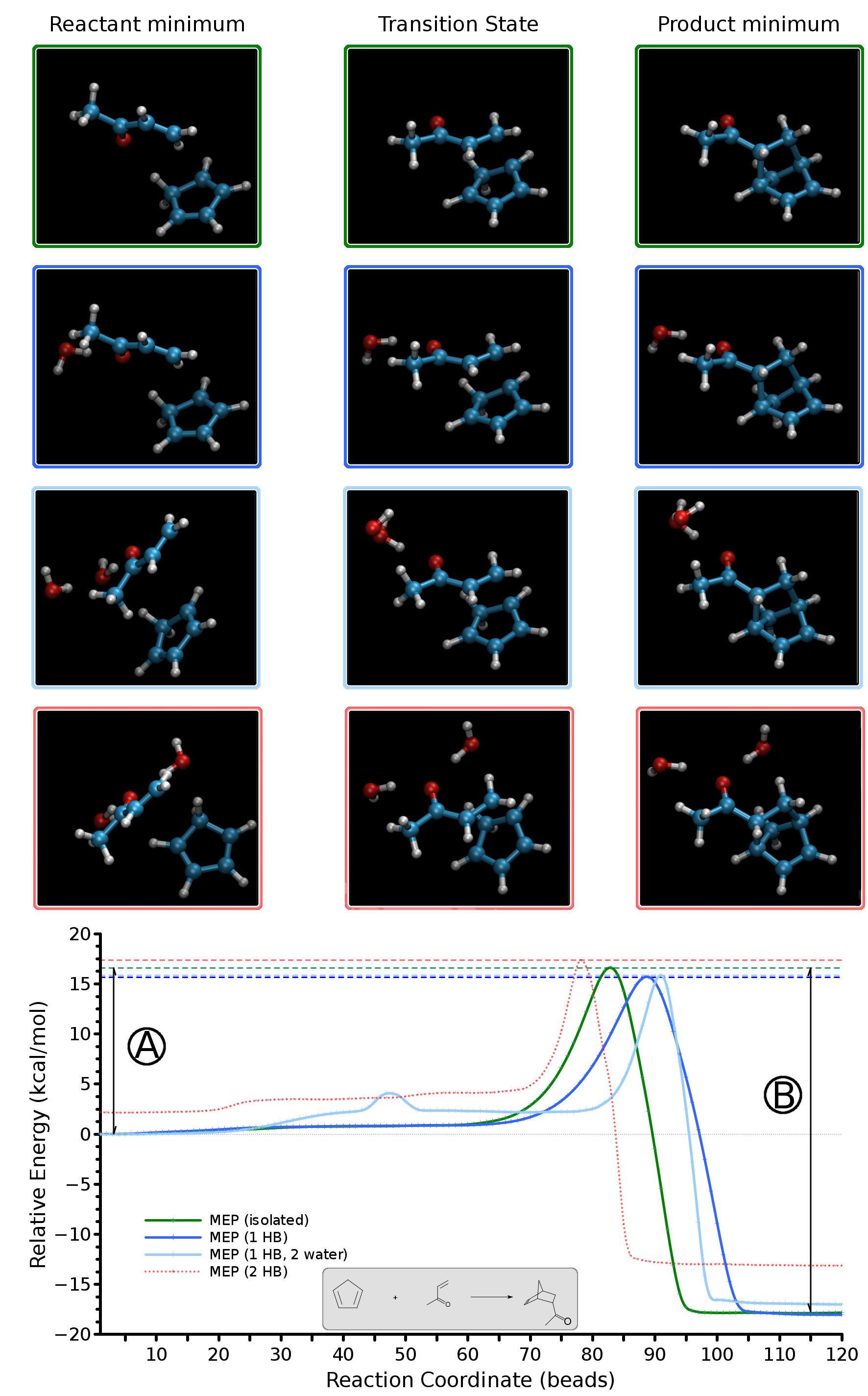}
\caption{The minimum energy paths of the DA reaction of
 cyclopentadiene and butenone obtained by the string method.
 The results are shown for the cases without water (green),
 with one water molecule (dark blue) and with two water molecules in a chain (light blue).
 The path for the case of with two water forming a double hydrogen bond with the ketone in plotted with a dotted red line. 
 Zero energy is set to that of the reactant minimum of each molecular cluster.
 The respective molecular structures of the reactant, transition,
 and product states are also given. The activation energy of forward reaction (\textcircled{A}): 16.6 kcal/mol (without water), 15.7 kcal/mol (with one water molecule) and 15.8 kcal/mol (with two water molecules). The activation energy of backward reaction (\textcircled{B}): 34.5 kcal/mol (without water), 33.7 kcal/mol (with one water molecule) and 32.8 kcal/mol (with two water molecules).}
\label{fig:bar-rest}
\end{figure}

In Figure \ref{fig:bar-rest}, the potential energy profiles
 along with the structures of stationary states
 are shown.
We find that the activation energy of the forward reaction
 without water molecule is 16.9 kcal/mol, but
 it is lowered by 0.9 kcal/mol when water molecule is added.
There are two possible conformations when adding more
 water molecules, i.e.
 the doubly hydrogen bonded conformation
 or the hydrogen bonded chain conformation
 (see Figure \ref{fig:bar-rest}).
According to our calculation, the latter is more stable.
 It is favoured for only one water molecule to be directly
 hydrogen bonded to butenone.
Moreover, the hydrogen bonded chain conformation
 has a lower activation energy than the doubly hydrogen
 bonded conformation.
However, the activation energy of the cluster system with two water molecules in the hydrogen bonded chain conformation is lower by 0.8 kcal/mol compared to that without water, almost the same that one water molecule cluster. The reduction of activation energy are qualitatively consistent with the experimental finding that the DA reaction is accelerated in water.
It is clear that the hydrogen bond plays the important
 part in the acceleration mechanism.
However, small increment of water molecules does not modify the acceleration rate,
and these values are smaller than
 the experimental estimate of the free energy
 barrier in liquid water \cite{dial-MKV-water-exp-1,dial-MKV-water-exp-2},
 where a change of 4 kcal/mol is found. 
Therefore, for more quantitative analysis, \textit{ab initio} studies
 with higher accuracy, and inclusion of solvent effects either by
 implicit or explicit treatment of water molecules, etc.
 should be useful for a future work.


\section{Conclusions}

Unification of PIMD and SMASH codes has enabled
 the hierarchical parallelization of \textit{ab initio} simulations
 with respect to replicas and forces.
As mentioned in the Introduction, this technique is open to a broad
 class of \textit{ab initio} simulations, such as quantum statistics,
 statistical sampling, nonadiabatic dynamics, reaction paths 
 or free energy exploration methods.
The parallel performance
 has been tested on the JAEA supercomputer with 2048 cores 
for two cases,
 i.e.\ path integral MD simulations of water molecules
 with various system sizes and reaction path calculation
 of the cyclopentadiene-butenone DA reaction by the string method.
The key measure in the feasibility of these \textit{ab initio}
 simulations is the average time per step.
This index
 drops as the number of cores are larger, and
 there is a minimum value of this measure reachable for
 a given system size.
More critically, there is an upper limit in
 the system size for a required value of time per step.
In the present case, this limitation comes from the
 deterioration of parallelization, in particular,
 the relative speed of the BO energy calculation
 shown in Figure \ref{fig:perf}b.
The strong scaling of large systems
 should be addressed in future
 development of \textit{ab initio} simulations.
The communication between replica-parallel is not
 very important in the case of this study.
In the hierarchical parallelization, there is a
 room of optimal choice of the replica- and
 force-parallel numbers.
%
The data obtained from this work should be useful not only for the application studies using multiple replica simulation but also for a further development of parallel codes and algorithms to be expected in the future.
Some improvements in the code are planning in a near future, such as the introduction of the hybrid parallelization MPI/OpenMP or the proper parallelization of coordination and velocities updating, to improve the shown efficiency.
The hierarchically parallelized code has been used
 to study the DA reaction of cyclopentadiene and butenone
 by the \textit{ab initio} string method.
The MEP is calculated in the absence of water, and
 in the presence of one and two water molecules hydrogen
 bonded to butenone.
The results indicated a lowering of the energy barrier
 in the presence of water qualitatively consistent with
 the experimental finding that the reaction rate is higher in liquid water than
 in the organic solvents.
For a more quantitative analysis including hydrophobic effects,
 \textit{ab initio} calculations with improved accuracy and taking
 account of the solvent effects should be indispensable.
%


\section*{Acknowledgement}
This work was supported by Theoretical and Computational Chemistry
Initiative (TCCI) and Computational Materials Science Initiative
(CMSI) in the Strategic Programs for Innovative Research, MEXT, Japan
(Project ID: hp150231), and JSPS KAKENHI Grant Number 25730079.
We thank the priority use of JAEA BX900 supercomputer
 in Center for Computational Science and E-systems at Tokai
 Research Establishment, JAEA.\\
We are grateful to Dr. A. Malins in JAEA for proofreading
 the manuscript of this paper.


\section*{References}

\bibliography{mybibfile}


\newpage

\begin{table}[p]
\centering
\caption{Average time per step in the path integral MD
 simulation of eight water molecules with
 various number of beads.
 Four cores are used per bead.
 All the beads are calculated in parallel.}
{ \small \begin{tabular}{l c c c c c c} \hline\hline
 Number of cores & 8 & 16 & 32 & 64& 128 & 256 \\
 Number of beads & 2 & 4 & 8 & 16 & 32 & 64 \\ \hline
 Time average [s] & 20.6 & 20.7 & 21.4 & 22.1 & 22.8 & 23.9  \\ \hline
\end{tabular}}
\label{table:pimd-perf1}
\end{table}
\begin{table}[p]
\centering
\caption{Relative speed in the path integral MD simulation
 of eight water molecules using various number of cores. Four cores are used per bead.}
{ \small \begin{tabular}{l c c c c c c} \hline\hline 
 Number of cores & 8 & 16 & 32 & 64& 128 & 256 \\
 Number of beads & 64 & 64 & 64 & 64 & 64 & 64\\ \hline
 Relative speed & 1 & 2.0 & 4.0 & 7.8 & 15.2 & 28.87 \\ \hline
\end{tabular}}
\label{table:pimd-perf2}
\end{table}
\newpage
\begin{table}[p]
\centering
\caption{ Transition state stabilization energy in water with respect to the gas phase reaction for DA reaction of cyclopentadiene and butenone of different works. Energies in kcal/mol. Free energies at $25^{\circ}$C and 1 atm. }
{ \small \begin{tabular}{r c c l}
 Method & $\Delta\Delta U^{\ddagger}$ & $\Delta\Delta G^{\ddagger}$ & Reference \\ \hline
 B3LYP/6-31G** 	& -0.9$^{a}$ 	& 		& This work \\
 B3LYP/6-31G** 	& -0.8$^{b}$ 	& 		& This work \\
 HF/6-31G* 	& -2.23$^{a}$ 	& 		& Ref\cite{dial-MVK-water-1} \\
 PDDG/PM3/MM/MC & 		& -3.5$^{c}$ 	& Ref\cite{dial-MVK-water-5} \\
 AM1/TIP3P 	& 		& -3.5$^{c}$ 	& Ref\cite{dial-MVK-water-4} \\
 AM1-OPLS-CM1A 	& 		& -2.8$^{c}$ 	& Ref\cite{dial-MVK-water-3} \\
 Experimental 	& 		& -3.8$^{d}$	& Ref\cite{dial-MKV-water-exp-1} \\
 Experimental 	& 		& -2.4$^{e}$	& Ref\cite{dial-MKV-water-exp-2} \\ \hline
\end{tabular}}
\\ {\footnotesize {\raggedright $^{a}$ With 1 water molecule. \\ $^{b}$ With 2 water molecules. \\  $^{c}$ Solvated in explicit water. \\ $^{d}$ Relative to isooctane at $20^{\circ}$C. \\ $^{e}$ Relative to 1-propanol. \\ }}

\label{table:dial-values}
\end{table}



\end{document}